\documentclass[pdftex,twocolumn,epjc3]{svjour3}          

\RequirePackage[T1]{fontenc}

\smartqed  
\RequirePackage{multirow}
\RequirePackage{graphicx}
\RequirePackage{mathptmx}      
\RequirePackage{flushend}
\RequirePackage[numbers,sort&compress]{natbib}
\RequirePackage[colorlinks,citecolor=blue,urlcolor=blue,linkcolor=blue]{hyperref}

\usepackage{graphicx}  
\usepackage{dcolumn}   
\usepackage{bm}        
\usepackage{amssymb}   
\usepackage{feynmf}
\usepackage{hyperref}
\usepackage{slashed}
\usepackage{multirow}
\usepackage{color}
\usepackage{array}
\usepackage{subcaption}
\usepackage{amsmath}
\usepackage{amssymb}
\usepackage{booktabs}
\usepackage{multirow}

\begin{document}

\title{Secondary Vertex Finding in Jets with Neural Networks}

\author{Jonathan Shlomi\thanksref{addr1},
	Sanmay Ganguly\thanksref{addr1},
	Eilam Gross\thanksref{addr1},
	Kyle Cranmer\thanksref{addr2}
	Yaron Lipman\thanksref{addr1},
	Hadar Serviansky\thanksref{addr1},
	Haggai Maron\thanksref{addr3},
	Nimrod Segol\thanksref{addr1},
}

\institute{Weizmann Institute Of Science, Israel\label{addr1}
	\and
	NYU\label{addr2}
	\and
	NVIDIA Research\label{addr3}
}

\date{Received: date / Accepted: date}

\maketitle

\begin{abstract}
	Jet classification is an important ingredient in measurements and searches for new physics at particle coliders, and secondary vertex reconstruction is a key intermediate step in building powerful jet classifiers.
	We use a neural network to perform vertex finding inside jets in order to improve the classification performance, with a focus on separation of bottom vs. charm flavor tagging.
	We implement a novel, universal set-to-graph model, which takes into account information from all tracks in a jet to determine if pairs of tracks originated from a common vertex. We explore different performance metrics and find our method to outperform traditional approaches in  accurate secondary vertex reconstruction. We also find that improved vertex finding leads to a significant improvement in jet classification performance.
\end{abstract}

\section{Introduction}

\begin{figure}[t]
	\begin{center}	
		\includegraphics[]{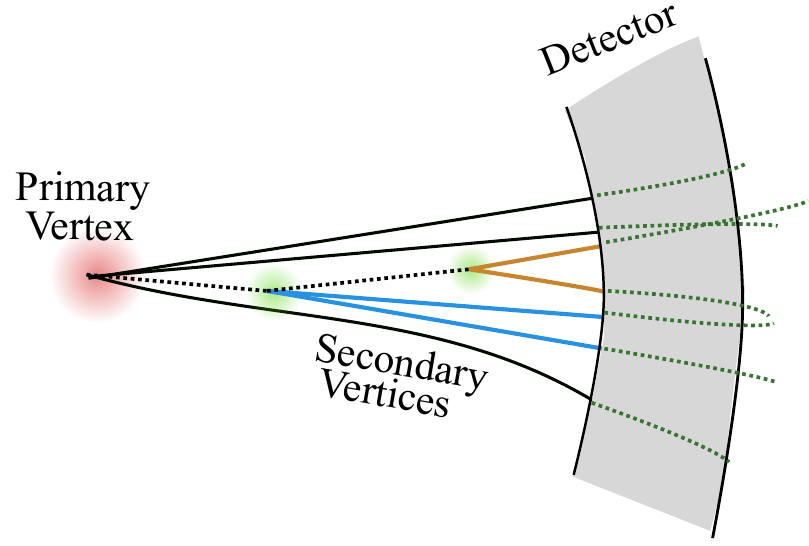}
		\caption{Illustration of a jet with secondary decay vertices. In order to identify the flavor of the jet, vertex reconstruction aims to group together the tracks measured in the detector based on their point of origin.}
		\label{fig:explain_vertexing}
	\end{center}
\end{figure}

Identifying jets containing bottom and charm hadrons and separating them from jets that originate from lighter quarks, is a critical task in the LHC physics program, referred to as "flavor tagging". Bottom and charm jets are characterized by the presence of secondary decays "inside" the jet - the bottom and charm hadrons will decay several millimeters past the primary interaction point (primary vertex), and only stable outgoing particles will be measured by the detector. Figure~\ref{fig:explain_vertexing} illustrates a typical bottom jet decay, with two consecutive displaced vertices from a bottom decay (blue lines) and charm decay (yellow lines).

Existing flavor tagging algorithms use a combination of low-level variables (the charged particle tracks, reconstructed secondary vertices), and high-level features engineered by experts as input to neural networks of various architectures in order to perform jet flavor classification~\cite{Guest:2016iqz}. 

Vertex reconstruction can be separated into two tasks, \emph{vertex finding}, and  \emph{vertex fitting}~\cite{Strandlie:2010zz}. Vertex finding refers to the task of partitioning the set of tracks, and vertex fitting refers to estimating the vertex positions given each sub-set of tracks. Existing algorithms typically use an iterative procedure of  finding and fitting to perform both tasks together. We focus on using a neural network for vertex finding only. Vertex finding is a challenging task because of two factors: 

\begin{itemize}
	\item  Secondary vertices can be in close proximity to the primary vertex, and to each other, within the measurement resolution of the track trajectories.  
	\item The charged particle multiplicity in each individual vertex is low, typically between 1 and 5 tracks.
\end{itemize}
Vertex reconstruction is in essence an inverse problem of a complicated noisy (forward) function:
\begin{equation}
	\text{Particle Decay} \rightarrow \text{Particle Measurement in Detector}
\end{equation}

Neural networks can find a model for this inverse problem without expert intervention by using supervised learning, i.e., by providing many examples of the forward process, which can be provided by simulations.
They can also be easily optimized by retraining without expert intervention. Particle colliders may have different modes of operation during their lifetime, such as the LHC increasing its collision energy over the years. Different data taking conditions require re-optimizing reconstruction algorithms, and neural networks provide a simple way to perform that re-optimization.

Since the set of tracks to be partitioned has no inherent order, we use an equivariant\footnote{If $x$ is an $n\times d$ tensor, and $\sigma$ is a permutation on $n$ elements, then a layer $L$ is called equivariant if $L(\sigma x)=\sigma L(x)$ and invariant if  $L(\sigma x)= L(x)$} neural network architecture. We show in this paper that this constraint on the model results in better performance.

We first describe the dataset on which we test our proposed algorithm in Section~\ref{sec:dataset}. The model architecture and the baseline algorithms
are described in Section~\ref{sec:methods}. Section~\ref{sec:defineperf} discusses the performance metrics defined for vertex finding. Section~\ref{sec:ftag_impact} describes how the impact of vertex finding on jet classification was assessed, and the results are presented in Section~\ref{sec:results}. Conclusions are given in Section~\ref{sec:conclusions}.

\subsection{Background}

\paragraph{Standard vertex reconstruction algorithms.}

Existing vertex reconstruction techniques are based on the geometry of the tracks, or a combination of the geometry and constraints that are configured by hand to match a specific particle decay pattern~\cite{Piacquadio_2008}. In order to handle finding and fitting multiple vertices, a standard algorithm is adaptive vertex reconstruction (AVR)~\cite{Strandlie:2010zz,Waltenberger:2011zz,Waltenberger:1166320}. The basic concept of AVR is to perform a least squares fit of the vertex position given all the tracks, then remove less compatible tracks from the fit, and refit those tracks again to more vertices. This repeats until no tracks are left. AVR can be used to first fit the primary vertex with special considerations for its unique properties, and subsequently fit secondary vertices. In this paper it is used as a general multi-vertex fitter, applied only to tracks associated to a single jet.

\paragraph{Deep learning on sets and graphs.} Following the successful application of deep learning to images \cite{lecun1998gradient, krizhevsky2012imagenet}, there is an ongoing research effort aimed at applying deep learning to other data structures such as unordered sets \cite{zaheer2017deep, qi2017pointnet,maron2020learning} and graphs \cite{Bruna2013, kipf,Gilmer2017,maron2018invariant}. Typical learning tasks for such domains are point-cloud classification for sets, or molecule property prediction, for graphs. A challenge in both scenarios stems from the arbitrary order of the elements in the set or the nodes in the graph. 
Fully connected, convolutional and recurrent networks do not have the correct inductive bias for learning tasks on unordered sets~\cite{battaglia2018relational}. They assume a fixed size or an ordering in the data. A popular design principle for networks that process such unordered data is constraining layers to be equivariant or invariant to the reordering operation. By using only equivariant layers the neural networks is constrained to represent only equivariant functions. 

Recently, the Set2Graph (S2G) model \cite{serviansky2020set2graph} was proposed as a simple, equivariant model for learning tasks in which the input is an arbitrarily ordered set of $n$ elements and the output is an $n\times n$ matrix that represents their pairwise relations. The S2G model was proved to be universal, meaning it can approximate any equivariant function from a set to a graph. We use this model in this paper.    

\paragraph{Deep learning for particle physics.} Neural networks that operate on sets have been used recently in a number of  particle physics applications~\cite{Shlomi_2020}. The data structure of an unordered set is a natural description for most particle physics reconstruction tasks, and recent progress in the field of graph neural networks~\cite{battaglia2018relational} has prompted many new applications. For the problem of track reconstruction, a graph neural network was used to classify the paths between adjacent detector "hits"~\cite{exatrkx,ju2020graph}. This is a similar application to vertex finding since the end result must be a partition of the set of hits to different tracks. Other applications of graph neural networks to partitioning sets of objects include particle reconstruction in calorimeters and liquid argon time projection chambers ~\cite{Kieseler_2020,drielsma2020clustering,Di_Bello_2021,Pata_2021}. Direct jet classification has also been proposed with a few different variants of message passing networks~\cite{JEDI,Qu:2019gqs, henrionneural, Komiske_2019,Moreno_2020,bernreuther2020casting,Mikuni_2020,guo2020boosted}.

\section{Data}
\label{sec:dataset}

We test the proposed algorithm on a simulated dataset\footnote{The dataset and code used in this paper are available at \url{https://zenodo.org/record/4044628} and \url{https://github.com/jshlomi/SetToGraphPaper}.}.
The dataset consists of jets sampled from  $pp\rightarrow t\bar{t}$ events at $\sqrt{s}=14$ TeV. The events are generated with {\sc pythia8}~\cite{Sj_strand_2015} and a basic detector simulation is performed with  {\sc delphes} ~\cite{de_Favereau_2014}, emulating a detector similar to ATLAS~\cite{Collaboration_2008}.
charged particle tracks are represented by 6 perigee parameters ($d_{0}$, $z_{0}$, $\phi$, $\text{cot}\theta$, $p_{T}$, $q$) and their covariance matrix. Noise is added to the track perigee parameters with Gaussian smearing. The track parameters resolution depends on the transverse momentum $p_{T}$ and pseudorapidity $\eta$ of the track in a qualitatively similar way to the measurements reported in~\cite{Collaboration_2008}. The covariance matrix is diagonal in this simplified track smearing model---the smearing is done independently for each parameter with no correlated effects.

Jets are constructed from calorimeter energy deposits with the anti-$k_{\mathrm{T}}$ algorithm~\cite{Cacciari:2008gp} with a distance parameter of $R = 0.4$. Charged tracks are cone associated to jets with a  $\Delta R  < 0.4$ cone around the jet axis. The flavor labeling of jets (as bottom, charm or light) is done by matching weakly decaying bottom and charm hadrons to the jet with a $\Delta R$ cone of size 0.3. 

A basic jet selection is applied, requiring jets have $p_{T}$ > 20 GeV and $|\eta|$ < 2.5
The input to the vertex finding algorithms is the set of tracks associated to each jet, the jet $p_{T}$, $\eta$, $\phi$ and jet mass.

\paragraph{Dataset composition.}

\begin{figure}[h]
	\begin{center}
		
		\begin{subfigure}[b]{.49\columnwidth}
			\includegraphics[]{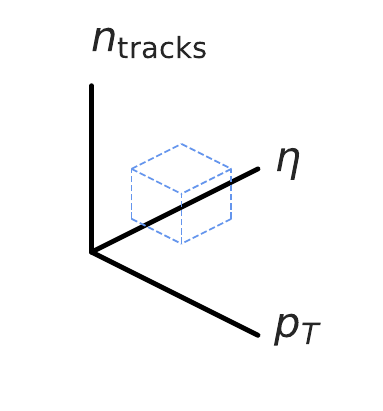}
			\caption{}
			\label{fig:datasample}
		\end{subfigure} 
		\begin{subfigure}[b]{.49\columnwidth}
			\includegraphics[]{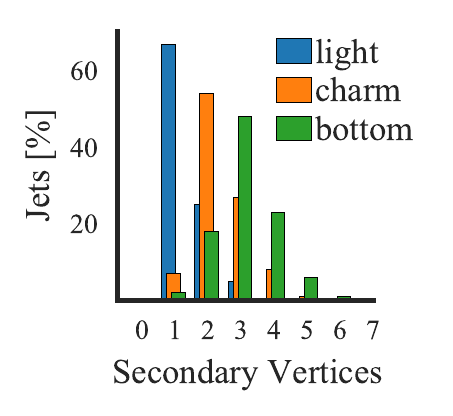}
			\caption{}
			\label{fig:npartitions}
		\end{subfigure} 
		\begin{subfigure}[b]{1\columnwidth}
			
			\includegraphics[]{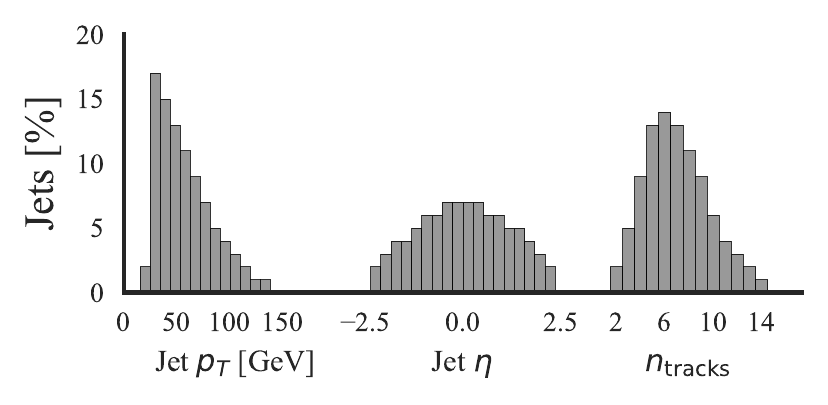}
			\caption{}
			\label{fig:datadist}
		\end{subfigure}
		\caption{(a) The dataset is composed by selecting equal numbers of jets from each flavor in each bin of $p_{T}$, $\eta$, and $n_\mathrm{tracks}$. (b) Distribution of the number of secondary vertices for the different jet flavors. (c) The resulting distribution of $p_{T}$, $\eta$, and $n_\mathrm{tracks}$ in the dataset.}
	\end{center}
\end{figure}

The properties of secondary vertices, such as their distance from the primary vertex, depend on the jet flavor but also on $p_{T}$, $\eta$, and number of tracks ($n_\mathrm{tracks}$). However, the distribution of those parameters is different for the different flavors, depending on the process used to generate the sample. The dataset is therefore built by sampling equal numbers of jets from each flavor in each $(p_{T},\eta,n_\mathrm{tracks})$ bin, as illustrated in Figure~\ref{fig:datasample}. For each bin, the flavor with the least amount of jets (usually c jets) in that bin determines the number of jets from the other flavors that are sampled. Figure~\ref{fig:npartitions} shows the resulting distribution of the number of vertices in each jet flavor, and Figure~\ref{fig:datadist} shows the distribution of  $p_{T}$, $\eta$, and $n_\mathrm{tracks}$ for all the flavors. The dataset is split into training (500k jets), validation, and testing datasets (100k jets each). 

\section{Vertex Finding Algorithms}
\label{sec:methods}

We compare 4 different algorithms.
\begin{itemize}
	\item Adaptive vertex reconstruction (AVR).
	\item  Set2Graph neural network.
	\item  Track pair (TP) classifier.
	\item  Recurrent neural network (RNN) model.
\end{itemize}

AVR serves as the baseline, and represents the existing vertex reconstruction algorithms.
The S2G model is our universal equivariant model. The TP and RNN algorithms are baseline neural networks that are similar to S2G but remove one of its important properties: The TP algorithm is not universal, while the RNN is not equivariant. The architectures of all models are described below.

\subsection{Adapative vertex reconstruction}

We use adaptive vertex reconstruction as  implemented in the RAVE software package~\cite{Waltenberger:2011zz}. This algorithm is a representative of existing (non neural network based) methods. The input to the algorithm is the set of tracks associated to the jet and their covariance matrix. The output is a set of vertices, and a set of track-to-vertex association weights. The algorithm can associate a track to more than one vertex. To convert this output into an unambiguous partition, each track is assigned to the vertex to which it has the highest weight. There are hyperparameters that control the iterative fitting or finding procedure such as cuts on the track-to-vertex weight for removing outliers, and these were scanned to find the most performant set of cuts based on the Rand index (defined in Section~\ref{sec:jetperf}). Additional details about the hyper-parameter optimization are given in~\ref{sec:raveoptim}.

\subsection{Set2Graph Neural Network}

\begin{figure}[h]
	\begin{center}	
		\includegraphics[]{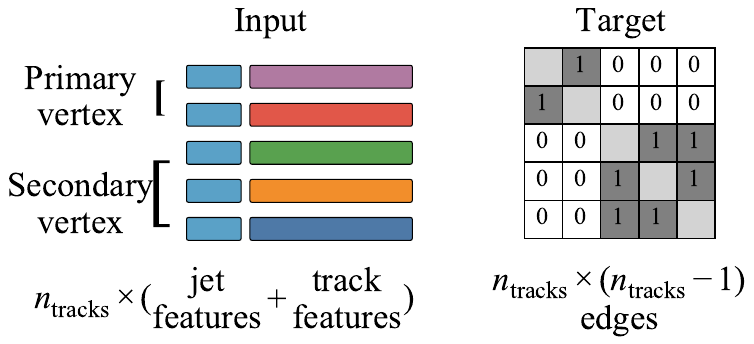}
		\caption{The input and training target for the neural network algorithms. For a jet with $n_\mathrm{tracks}$, the input is an array of $n_\mathrm{tracks} \times d_\mathrm{in}$ track and jet features (jet features are represented by the light blue boxes, track features by the colored boxes), and the target output is a binary classification label for each of the  $n_\mathrm{tracks}\times(n_\mathrm{tracks}-1)$ ordered pairs of tracks in the jet.}
		\label{fig:taskdefine}
	\end{center}
\end{figure}

For the neural network training, the vertex finding task is cast as an edge classification task, as illustrated in Figure~\ref{fig:taskdefine}. The input consists of the tracks associated to a jet, represented as an array of $n_\mathrm{tracks} \times d_\mathrm{in}$ matrix, with the $d_\mathrm{in}=10$ features composed of the 6 track perigee parameters and the jet feature vector (the jet features are duplicated for each track). The output is a binary label attached to each pair of tracks indicating whether they originated from the same position in space.

\begin{figure*}[t]
	\begin{center}	
		\includegraphics[]{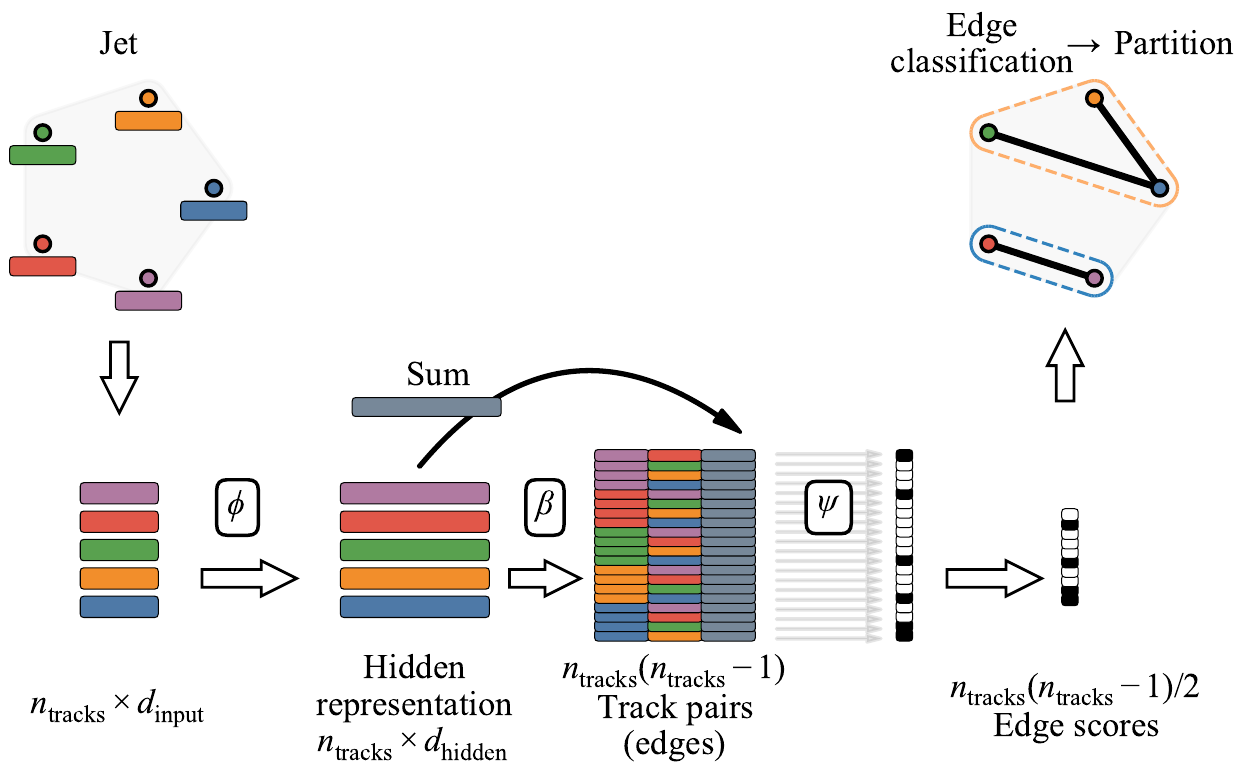}
		\caption{Partitioning a set of jet tracks using a neural network. A set-to-set component, $\phi$, creates a hidden representation of each track, with size $d_{\text{hidden}}$. A broadcasting layer $\beta$, then creates a representation for each directed edge (ordered pair of tracks in the jet) by combining the representation of the two tracks and the sum of all representations. An edge classifier $\psi$ then operates on the directed edges. This output is used for training the model (see the target definition in Figure~\ref{fig:taskdefine}). During inference the output of the edge classifier is symmetrized to produce an edge score. The edge scores are used to define the set partition by optimizing the partition score, as described in~\ref{sec:inference}}
		\label{fig:settograph}
	\end{center}
\end{figure*}

The S2G network is built as a composition of 3 modules, $\psi\circ\beta\circ \phi$: a set-to-set component, $\phi$, a broadcasting layer $\beta$ and a final edge classifier $\psi$. Here we give only a high level description of what each module does and its purpose, the specific model details are given in~\ref{sec:moremodel}. The model architecture is illustrated in Figure~\ref{fig:settograph}.

The set-to-set component $\phi$ takes as input the matrix of size $n_\mathrm{tracks} \times d_\mathrm{in}$. The output of $\phi$ is a hidden representation vector for each track, with size $n_\mathrm{tracks} \times d_{\mathrm{hidden}}$. $\phi$ is where information is exchanged between tracks and it is implemented as a deep sets~\cite{zaheer2017deep} network.

The broadcasting layer $\beta$ constructs a representation for each ordered pair of tracks (directed edge) using the output of $\phi$. The edge representation is simply a concatenation of the representations of the two tracks, with the sum of all track representations, resulting in an output of size $(n_\mathrm{track}(n_\mathrm{track}-1))\times 3 d_\mathrm{hidden}$.

The edge classifier $\psi$ is an MLP that operates on the edges to produce an edge score. This edge score is trained according to the target defined in Figure~\ref{fig:taskdefine}. During inference (after the training is complete) the edge scores are symmetrized, so for an unordered track pair the edge score $s_{ij}$ is:
\begin{equation}\label{eq:edgescore}
	s_{ij} = \sigma( \frac{1}{2} \left( \psi(\text{track}_{i}, \text{track}_{j})+\psi(\text{track}_{j}, \text{track}_{i}) \right) )
\end{equation}

Where $\sigma$ is the sigmoid function.

\subsection{Neural network baselines}
The neural network baselines are meant to check the importance of the properties of the S2G model. The models have a similair number of trainable parameters: 0.46M for S2G, 0.42M and 0.53M for TP and RNN respectively. They share the same architecture of $\psi \circ \beta \circ \phi$ as the S2G model, with some components replaced as described below. Their properties are summarized in Table~\ref{tab:modelcompare}.

\begin{table}[]
	\centering
	\begin{tabular}{l|c|c|c|c}
		Model &  \begin{tabular}{@{}c@{}}Equivariant/ \\ Universal\end{tabular} & MFLOPS & Parameters & \begin{tabular}{@{}c@{}}Inference\\ time [ms]\end{tabular} \\
		\hline
		Set2Graph & \checkmark \checkmark & 7.7  & 4.6M & 5.5 \\
		Track Pair & \checkmark  $\text{\sffamily X}$ & 6.9 &  4.5M  & 2.9 \\
		RNN & $\text{\sffamily X}$ \checkmark & 9.1 &  5.3M & 23.4
	\end{tabular}
	\caption{Comparison of the neural network models. The inference time and FLOPS are measured per single jet with 14 tracks. FLOPS were estimated with~\cite{ptflops}.}
	\label{tab:modelcompare}
\end{table}

\begin{figure}
	\centering
	\includegraphics[]{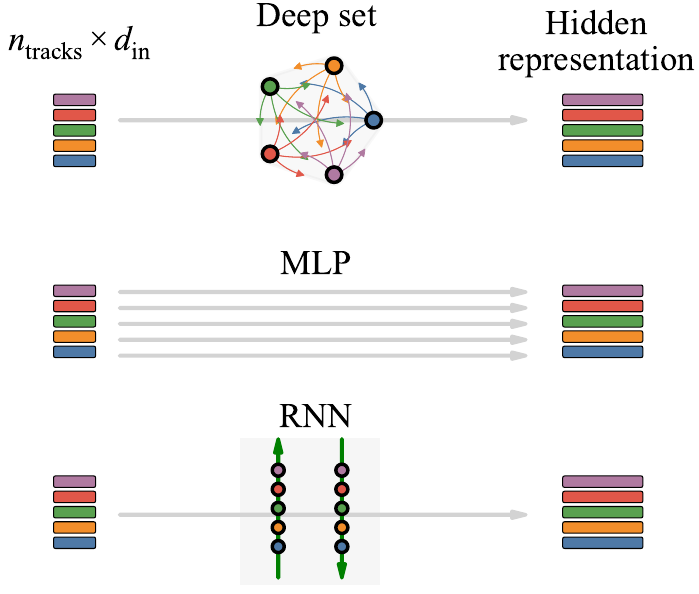} 
	\caption{The deep set module $\phi$ in the S2G model (top) creates the track hidden representation based on information exchange between the tracks in the jet. The TP classifier (center) however, creates the hidden representation with an MLP, which operates on each track individually. The RNN model (bottom) creates the hidden representation with a bi-directional GRU, which means the output depends on the order in which the tracks are sorted.}
	\label{fig:set2gvsmlp}
\end{figure}

The TP classifier is not a universal model. It will allow us to quantify the contribution of the information exchange between tracks to the overall vertex finding performance. 
As illustrated in Figure~\ref{fig:set2gvsmlp}, the hidden representation created for each track by the deep set module is conditional on the other tracks in the jet. We expect that for the task of vertex finding, being aware of all tracks is important, as the probability of a track pair being connected is conditional on the presence or lack of additional tracks nearby. 

The TP classifier checks this assumption about the data. If the probability of each track pair is conditional only on the properties of the track pair, this algorithm will perform as well as the S2G model. It is still expected to perform reasonably well, as it can still learn to join together tracks based on their geometry alone.

The deep set based $\phi$ layer is replaced by an MLP applied to each track in the jet (independently from the other tracks) to produce some hidden vector representation of that track. While a deep set has been proven to be universal (can approximate any function from sets to sets) \cite{segol2019universal} applying elementwise MLP is not universal for permutation equivariant functions.

Additionally, the broadcasting layer $\beta$ does not use the sum of the track hidden representations. The $\psi$ network operates only on the pair of track hidden representations.
Therefore in the TP classifier there is no information exchange between the track pairs---each track pair is classified independently.

In the RNN model the $\phi$ deep set component is replaced by a stack of bi-directional GRU layers~\cite{cho2014learning}. Each GRU layer processes the sequence of track representations, sorted by the track transverse momentum. The layer output is a concatenation of the sequence of hidden representations from both directional passes of the GRU, therefore each track hidden representation still contains information from all other tracks in the jet. This model can theoretically learn any function that the S2G model can, but its architecture is not equivarient. This model will show if the equivariance is a useful inductive bias for this task. Additionally, the sequential nature of the RNN leads to a slower inference time compared to the S2G and TP models (see Table~\ref{tab:modelcompare}).

\subsection{Inference}
\label{sec:inference}
The network output needs to be converted into a cluster assignment for the tracks. If an edge tracks $i\rightarrow j$ is connected, and track $j$ is connected to track $k$, then the edge between $i\rightarrow k$ must also be connected, regardless of its edges score. This could lead to a situation where many edges with low edge scores are artificially connected. Therefore we utilize the partition score optimization algorithm proposed by the authors of~\cite{drielsma2020clustering}. Track pairs whose score (eq. \ref{eq:edgescore}) is above a threshold of 0.5 are considered in sequence of decreasing score, and are "connected" only if their addition decreases the partition score:
\begin{equation}
	\textrm{Partition score} = \sum \delta_{ij}\ln(s_{ij})+(1-\delta_{ij})\ln(1-s_{ij})
\end{equation}

where $\delta_{ij}$ is 1 if $ \text{track}_{i}$ and $ \text{track}_{j}$ are assigned to the same cluster. In other words, if the connection of two tracks leads to an indirect connection between tracks with low edge scores, the connection is rejected.

\subsection{Training procedure and Loss function}

We train the network $f$ to perform edge predictions, i.e., predicting the probability of each pair of input tracks to originate from the same vertex. For a jet with $n_\mathrm{tracks}$ we therefore predict $n_\mathrm{tracks} (n_\mathrm{tracks}-1)$  edge scores. We train the network $f$ with the edge predictions before the symmetrization step, which results in  $n_\mathrm{tracks} (n_\mathrm{tracks}-1)/2$ edge scores. 

In terms of edge classification, it is import to balance the false positive and negative rates. We initially trained the network with a standard binary cross entropy (BCE) loss function:
\begin{equation}\label{eq:bceloss}
	\text{BCE}= \sum_\mathrm{edges} -y_\mathrm{edge} \ln(\hat{y}_\mathrm{edge}) -(1-y_\mathrm{edge}) \ln(1-\hat{y}_\mathrm{edge}) 
\end{equation}
where $\hat{y}_\mathrm{edge}$ is the edge predicted value, between 0 and 1, and $y_\mathrm{edge}$ is the truth edge label (0 or 1). The sum is over all edges in a single jet.

Training with BCE loss function resulted in a high number of false negatives. We therefore introduced a loss function based on the $F_\beta$ score, defined as:
\begin{equation}\label{eq:f1score}
	F_\beta= \frac{(1+\beta^2)\cdot \text{TP}}{ 
		(1+\beta^2)\cdot \text{TP}+ \text{FP}+\beta^2\cdot \text{FN}  } 
\end{equation}

with TP, FP, FN the true positives, false positives and false negatives respectively. The $F_\beta$ score is not differentiable. Quantities such as \emph{true positives} are defined by functions that contain non differentiable conditions, for example:
\begin{equation}\label{eq:tp}
	\textrm{true positives}\equiv \sum_{\mathrm{edges}}( \hat{y}_{\mathrm{edge}} > \textrm{threshold} ) y_\mathrm{edge}
\end{equation}

To compute a differentiable $F_\beta$ loss, denoted as $F_\beta^*$ these quantities are approximated as differentiable functions:
\begin{equation}\label{eq:contvar}
	\begin{aligned}
		&\text{true positives}^*\equiv \sum_{\mathrm{edges}}\hat{y}_{\mathrm{edge}}\cdot y_{\mathrm{edge}} \\
		&\text{false positives}^*\equiv \sum_{\mathrm{edges}}\hat{y}_{\mathrm{edge}}\cdot(1- y_{\mathrm{edge}}) \\
		&\text{false negatives}^*\equiv \sum_{\mathrm{edges}}(1-\hat{y}_{\mathrm{edge}})\cdot y_{\mathrm{edge}}
	\end{aligned}
\end{equation}

However, training with the $F_\beta^*$ loss only was unstable. Given the random weight initialization of the network, the training would sometimes fail to converge. A combined loss of BCE and F1 was finally used:
\begin{equation}\label{eq:loss}
	\text{Loss}= \text{BCE}-\lambda \sum_{\mathrm{jets}}F_\beta^*
\end{equation}

$\lambda$ and $\beta$ are hyperparameters that control the balance between false negatives and false positives.

\section{Performance Metrics for Vertex Finding}
\label{sec:defineperf}
We quantify the vertex finding performance from 3 different perspectives: The entire jet, individual vertices and pairs of vertices.
The motivation for defining multiple metrics is that vertex finding is an intermediate step which is used for a number of other tasks related to event reconstruction. Therefore it is important to quantify the performance for a wide variety of jets with different kind of decay topologies.

\subsection{Overall Jet Performance}
\label{sec:jetperf}
For jets as a whole, we consider the adjusted Rand index (ARI)~\cite{ARI}. ARI is a measure of the similarity between two set partitions. For vertex finding where the ground truth is well defined, we can treat the ARI of a jet as a "score" that tells us how well our vertex finding algorithm reproduced the ground truth partition. ARI is a normalized form of the Rand index, defined as:

\begin{equation}\label{randindex}
	\mathrm{RI} = \frac{\text{number of correct edges}}{ \text{number of edges in the set} }
\end{equation}

Correct edges are edges whose label matches the label they have in the ground truth (true positives and true negatives).
The adjustment of the RI is done by normalizing relative to the expectation value or the RI:
\begin{equation}\label{adjustedrandindex}
	\mathrm{ARI}=\frac{\mathrm{RI}-\mathbb{E}[{\mathrm{RI}}]}{1-\mathbb{E}[\mathrm{RI}]}
\end{equation}

The expectation value of the RI is defined by a choice of a random clustering model. There are several models one can adopt, described in Ref~\cite{ARIassumptions}. In our case a suitable choice is the "one-sided" comparison, where the true vertex assignment is considered fixed, and the expectation value is computed assuming one draws a completely random vertex assignment for the algorithm prediction. The expression for the expectation value is therefore:

\begin{equation}\label{randindexepected}
	\mathbb{E}[{\mathrm{RI}}] = \frac{B_{N-1}}{B_{N}}\frac{\sum_{i} {g_{i} \choose 2} }{ {N \choose 2} }
	+ \left(1- \frac{B_{N-1}}{B_{N}} \right) \left( 1- \frac{\sum_{i} {g_{i} \choose 2} }{ {N \choose 2} } \right)
\end{equation}
where $N\equiv n_\mathrm{tracks}$, $B_{N}$ is the bell number (the number of possible partitions of a set with $N$ elements), the sum is over the $i$ vertices in the jet and $g_{i}$ is the number of tracks in the $i$-th vertex.

An ARI score of 1 means the algorithm found the correct cluster assignment, while 0 represents a cluster assignment that is as good as random guessing. We consider the ARI score in 3 categories: perfect (ARI of 1), intermediate (ARI between 0.5 and 1), and poor (ARI lower than 0.5).

\subsection{Vertices and Vertex-Pairs Performance}
\label{sec:performancemetrics}

\begin{figure*}[]
	\begin{center}	
		\includegraphics[]{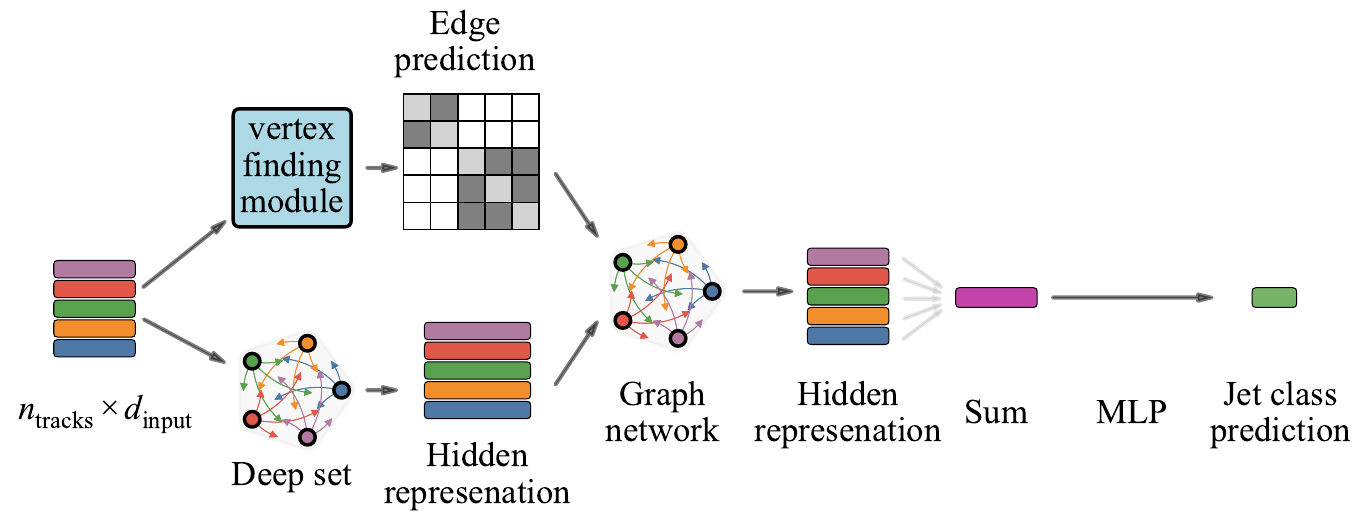}
		\caption{Jet classification model. The vertex finding module contains either one of the neural network models described in section~\ref{sec:methods}, or the predictions produced by the baseline AVR algorithms, pre-computed on the training dataset. If a pre-trained network in used in the vertex finding module, its weights are frozen during the training of the jet classifier.}
		\label{fig:jetclassification}
	\end{center}
\end{figure*}

\begin{figure}
	\centering
	\includegraphics[]{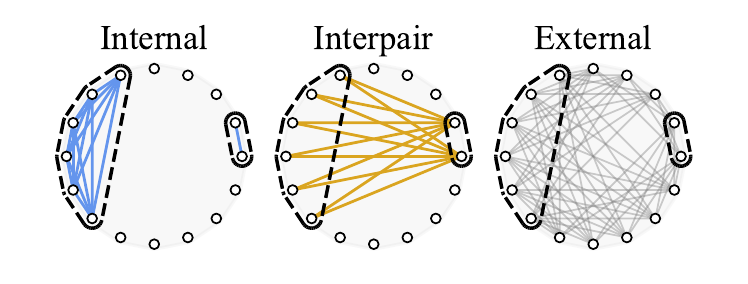} 
	\caption{Definition of internal, interpair and external edges for a pair of vertices.}
	\label{fig:explain_vertexpair}
\end{figure}

Instead of looking at an entire jet, we can consider subsets of the jet---individual vertices and all possible vertex pairs. We distinguish between \emph{internal}, \emph{external}, and \emph{inter-pair} edges. Figure~\ref{fig:explain_vertexpair} illustrates the definition. Internal edges connect tracks inside a vertex, Interpair edges connect tracks in one vertex to tracks in the other vertex (this definition is only relevant for vertex pairs), and external edges connect tracks from the vertex/vertex pair to other tracks in the jet. Note that "external edges" refers to edges that are connected only at one end to one of the tracks in the subset under consideration (vertex or vertex pair)---not to all edges that are external to the subset.
Considering a specific vertex, or a pair of vertices, we can compute separately the accuracy for each type of edge:

\begin{equation}\label{eq:accuracy}
	\text{Accuracy}_{\text{edge type}}= \frac{\text{correct edges}}{ \text{number of edges of that type}} 
\end{equation}
where for internal edges, \emph{correct edges} are those predicted to be connected by the algorithm, and for the other types, correct edges are those predicted to be disconnected.

\begin{table*}[ht]\centering
	\begin{tabular}{llp{5mm}p{4mm}p{7mm}|p{10mm}p{10mm}p{10mm}|p{10mm}p{10mm}p{10mm}p{10mm}p{10mm}}
		\toprule
		\multirow{2}[3]{*}{} & 
		\multirow{2}[3]{*}{Algorithm} & 
		\multicolumn{3}{c}{Jet} & 
		\multicolumn{3}{c}{Vertex} &
		\multicolumn{4}{c}{Vertex-Pair} \\
		\cmidrule(lr){3-5} \cmidrule(lr){6-8} \cmidrule(lr){9-13}
		& & F1 & RI & ARI & $\text{internal}$ & $\text{external}$ & $\text{combined}$ &  $\text{internal}_{1}$ &  $\text{internal}_{2}$ & $\text{interpair}$& $\text{external}$ &$\text{combined}$  \\
		\midrule
		
		\hline
		b jets & AVR & 0.56 & 0.61 & -0.01 & \textbf{0.91} & 0.51 & 0.46 & \textbf{0.59} & \textbf{0.90} & 0.54 & 0.58 & 0.18 \\
		& Track Pair & 0.62 & 0.74 & 0.32 & 0.86 & 0.71 & 0.60 & 0.55 & 0.87 & 0.72 & 0.74 & 0.29 \\
		& RNN & 0.59 & 0.75 & 0.37 & 0.79 & \textbf{0.77} & 0.60 & 0.48 & 0.84 & \textbf{0.78} & \textbf{0.80} & 0.27 \\
		& Set2Graph & \textbf{0.66} & \textbf{0.78} & \textbf{0.43} & 0.86 & 0.76 & \textbf{0.65} & 0.54 & 0.88 & \textbf{0.78} & 0.79 & \textbf{0.33} \\
		\hline
		c jets & AVR & 0.70 & 0.65 & 0.22 & \textbf{0.95} & 0.41 & 0.39 & \textbf{0.49} & \textbf{0.91} & 0.49 & 0.66 & 0.14 \\
		& Track Pair & 0.74 & 0.73 & 0.40 & 0.92 & 0.58 & 0.52 & 0.47 & 0.88 & 0.65 & 0.76 & 0.24 \\
		& RNN & 0.71 & 0.72 & 0.40 & 0.86 & \textbf{0.60} & 0.50 & 0.39 & 0.85 & 0.65 & 0.77 & 0.19 \\
		& Set2Graph & \textbf{0.75} & \textbf{0.75} & \textbf{0.45} & 0.94 & \textbf{0.60} & \textbf{0.56} & 0.47 & \textbf{0.91} & \textbf{0.67} & \textbf{0.78} & \textbf{0.26} \\
		\hline
		light jets & AVR & \textbf{0.97} & \textbf{0.96} & 0.93 & \textbf{0.99} & 0.89 & 0.88 & \textbf{0.33} & \textbf{0.98} & 0.73 & 0.89 & 0.14 \\
		& Track Pair & 0.96 & \textbf{0.96} & 0.93 & 0.97 & \textbf{0.93} & 0.90 & 0.32 & 0.97 & 0.87 & \textbf{0.95} & \textbf{0.26} \\
		& RNN & 0.93 & 0.92 & 0.87 & 0.93 & 0.90 & 0.84 & 0.25 & 0.94 & 0.82 & 0.93 & 0.18 \\
		& Set2Graph & \textbf{0.97} & \textbf{0.96} & \textbf{0.94} & 0.98 & \textbf{0.93} & \textbf{0.91} & 0.32 & \textbf{0.98} & \textbf{0.88} & \textbf{0.95} & \textbf{0.26} \\
		
		\bottomrule
	\end{tabular}
	\caption{Comparing vertex finding performance from three perspectives: jet, vertex and vertex-pair. See section~\ref{sec:defineperf} for the definitions of the various metrics. The mean for each metric, split by jet flavor, is shown for the S2G, AVR and TP algorithms. The S2G model outperforms or equals the other algorithms, maintaining the baseline AVR high accuracy for light jets with significant improvements for b and c jets.}
	\label{tab:vertexperformance}
\end{table*}

We can also multiply the different kinds of accuracies to compute an overall accuracy for the vertex/vertex-pair in question~\footnote{For vertices without one kind of edge (e.g vertex with 1 track and no internal edges) the accuracy for that type is set to 1}.

For individual vertices, we can evaluate the accuracy as a function of any vertex property we deem important, for example the number of tracks in the vertex. For vertex pairs, an important metric is the performance as a function of the distance between the two vertices. It is expected that as the distance between vertices decreases, accurate vertex finding becomes more difficult, and nearby vertices will be merged. The vertex pair performance metrics allow us to quantify that.

\section{Impact on Jet Classification}
\label{sec:ftag_impact}
In order to asses the impact of improved vertex finding on jet classification, we trained a classifier that took the edge classification prediction of the different algorithms as input, along with the tracks and jet features. The classifier predicts if the jet is a bottom, charm or light jet. The architecture for jet classification is illustrated in Figure~\ref{fig:jetclassification}. A vertex finding module (either AVR, or one of the neural network models) is used to produce an edge prediction for the input set of tracks, which is added to a hidden representation created by a deep set. The resulting graph is processed by a graph network~\cite{battaglia2018relational} and the resulting graph representation is classified by an MLP. Details about the architecture and training are given in~\ref{sec:jetclassificationappendix}. In this scenario, the edge predictions can be considered as a form of supervised attention for the jet classifier. The weights of the vertex finding module are frozen during training.

The baseline classification performance is given by training the same model with an untrained S2G vertex finding module. This baseline model has the ability to reach the same performance as the model with the pre-trained S2G network, as it is an identical network. However it is trained only with the classification objective, where both vertex finding module and the rest of the network are trained together. This baseline therefore shows if an unsupervised attention mechanism can reach similar classification performance, which would require it to identify the relevant features in the data without guidance.

\section{Results}
\label{sec:results}

\begin{figure}
	\begin{center}
		\includegraphics[]{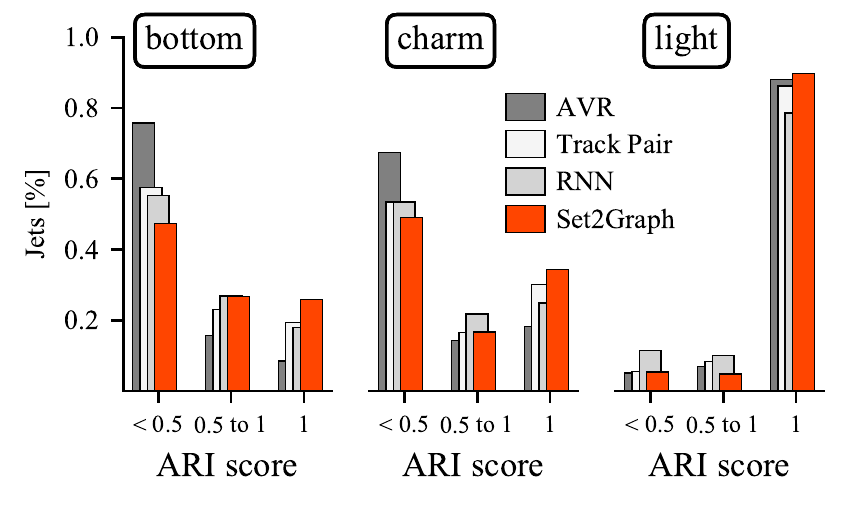}
	\end{center}
	\caption{ARI scores for the different flavors of jets. We consider 3 categories: \emph{Perfect}---jets with an ARI score of exactly 1, \emph{Intermediate}---a score between 0.5 and 1 and \emph{Poor}---scores below 0.5}
	\label{fig:ariperf}
\end{figure}

\begin{figure}
	\begin{center}
		\includegraphics[]{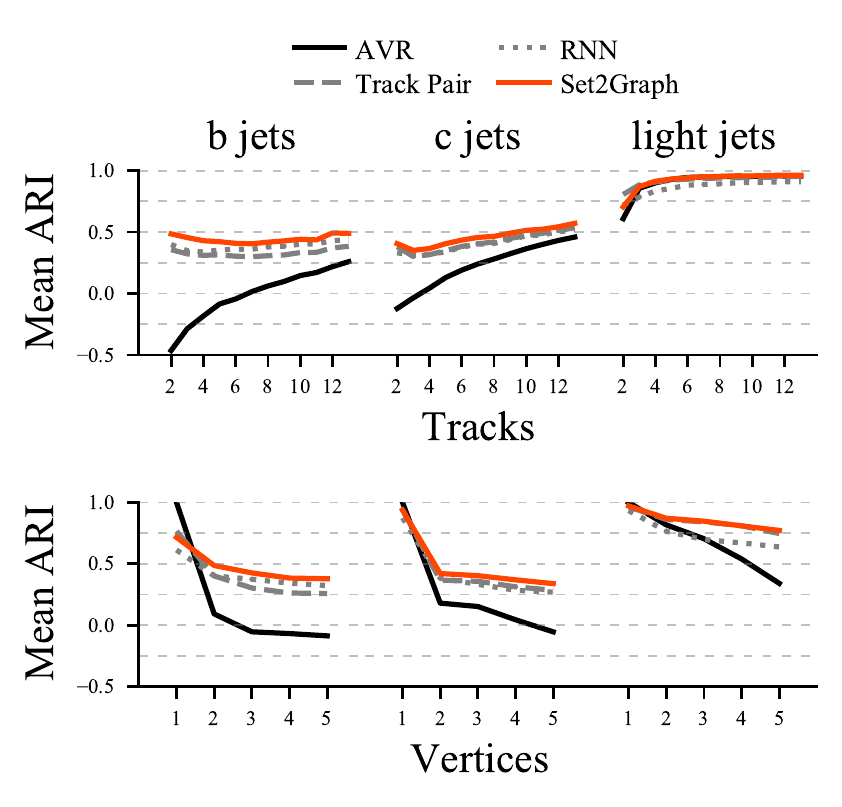}
	\end{center}
	\caption{Mean ARI scores for the different flavors of jets as a function of the jet properties.}
	\label{fig:ariperf_vs_vars}
\end{figure}

The vertex finding results are summarized in Table ~\ref{tab:vertexperformance}. The S2G model outperforms AVR in all jet performance metrics. The improvement is significant (about 20\% increase in ARI) for b and c jets, while for light jets the same high performance is maintained. The ARI distribution for the different flavors is shown in Figure~\ref{fig:ariperf} --- while there is still a substantial amount of poorly reconstructed jets (with ARI < 0.5) there are more than twice as many perfectly reconstructed b and c jets compared to AVR. In Figure~\ref{fig:ariperf_vs_vars} the mean ARI is shown as a function of both the number of tracks, and the number of vertices in the jet. For b jets, there is a very large improvement in jets with a small number of tracks, but the advantage over AVR is maintained across the entire range. The AVR algorithm outperforms S2G only in b and c jets which have only one vertex, which are very rare in the dataset.

\begin{figure}
	\begin{center}
		\includegraphics[]{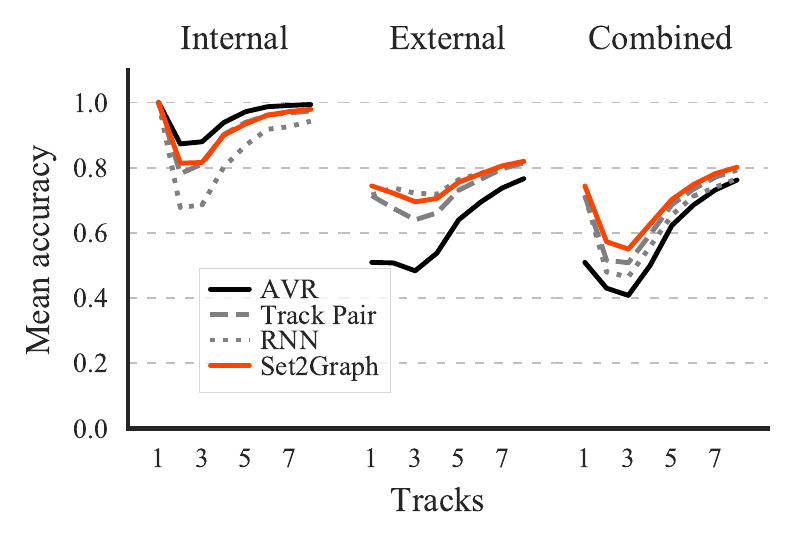}
	\end{center}
	\caption{Vertex performance as a function of the vertex size. Internal, external and combined accuracy are defined in Section~\ref{sec:performancemetrics}}
	\label{fig:vertex_performance}
\end{figure}

When considering vertex and vertex-pair metrics, for bottom and charm jets the mean internal accuracy  for S2G is within 1\% of the baseline, and a large increase (between 10 to 20\%) is achieved for external and inter-pair accuracy. Figure~\ref{fig:vertex_performance} shows the performance for vertices, as a function of vertex size (i.e., number of tracks in the vertex). The S2G algorithm maintains an advantage over the full range of vertex sizes. The S2G model has a similar internal accuracy to the baseline, but a 10\% increase in external accuracy for smaller vertices.

\begin{figure}
	\begin{center}
		\includegraphics[]{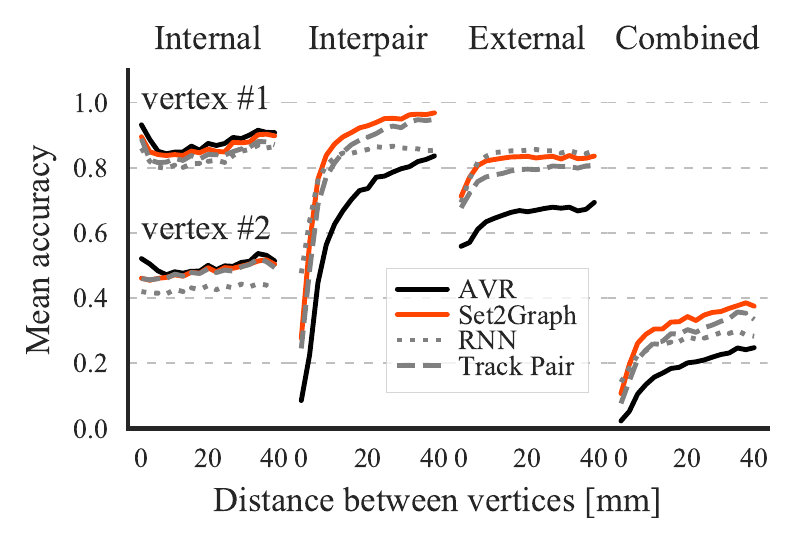}
	\end{center}
	\caption{Vertex pair accuracy as a function of distance between the vertices. The internal accuracy is shown for both smaller vertex (the vertex with fewer tracks, vertex \#1) and the larger vertex (vertex \#2).}
	\label{fig:vertex_pair_performance}
\end{figure}

Figure~\ref{fig:vertex_pair_performance} show the performance for vertex pairs, as a function of the distance between the vertices.  Again the S2G shows a promising ability to separate vertices even when the distance between them approaches 0. The performance increase  of about 10\% in combined accuracy comes from the improvement in interpair and external accuracy, i.e., less merging of vertices. 

\paragraph{Comparison to neural network baselines}
Both the TP and RNN algorithms have a lower ARI by about 5 to 10\% compared to the S2G model for b and c jets. S2G also outperforms both baselines in vertex and vertex-pair combined accuracy. From Figure~\ref{fig:ariperf} we can see that S2G has the highest percentage of perfectly reconstructed jets, and and Figures~\ref{fig:ariperf_vs_vars}, \ref{fig:vertex_performance} and \ref{fig:vertex_pair_performance} show that this advantage is maintained across the entire dataset.

\begin{table}[]
	\centering
	\begin{tabular}{l|c|c|c|c|c}
		\begin{tabular}{@{}c@{}}Vertex Finding \\ Module\end{tabular} & Accuracy & F1 
		& \begin{tabular}{@{}c@{}}b jets \\ F1\end{tabular}  &  \begin{tabular}{@{}c@{}}c jets \\ F1\end{tabular} & \begin{tabular}{@{}c@{}}light jets \\ F1\end{tabular}  \\
		\hline
		AVR & 0.50 & 0.49 & 0.62 & 0.44 & 0.40 \\
		Baseline & 0.57 & 0.56 & 0.67 & 0.40 & 0.60 \\
		Track Pair & 0.56 & 0.57 & 0.65 & \textbf{0.48} & 0.57 \\
		RNN & 0.62 & 0.60 & \textbf{0.74} & 0.37 & \textbf{0.69} \\
		Set2Graph & \textbf{0.63} & \textbf{0.62} & 0.72 & 0.44 & \textbf{0.69} \\
	\end{tabular}
	\caption{Jet flavor classification performance metrics. The model with pre-trained S2G vertex finding module outperforms the other algorithms in overall }
	\label{tab:classifierresults}
\end{table}

\paragraph{Impact on jet classification}
The results for jet classification are shown in Table~\ref{tab:classifierresults}. The pre-trained S2G classifier outperforms the AVR based classifier by over 10\% in terms of overall accuracy with the most significant gain coming from the increased rejection of light jets (an increase in light jet F1 from 40\% to 69\%). The neural network baseline with an S2G based vertexing module that is trained only towards the classification objective shows better performance than the AVR and Track Pair based algorithms. This indicates that the network is able to learn some important features of the data by itself. The RNN and S2G based models have similar performance, with the S2G model outperforming the RNN in particular in c jet identification.

\section{Conclusions}
\label{sec:conclusions}

We proposed training a neural network to perform vertex finding, using supervised learning. We found that it outperforms standard techniques for multiple performance metrics of vertex reconstruction, and shows promising increase in performance for nearby vertices.

We utilized the Set2Graph model, a simple equivariant and universal model of functions from sets to graphs. We showed that the model's universality and equivariance were both important. The universality was needed to properly learn the vertex finding task, by taking into account information from all tracks in the jet. Equivariance was a useful inductive bias, resulting in better performance compared to recurrent neural network which could in theory learn the same function as the S2G model. 
We evaluated the impact of the improved accuracy in vertex reconstruction on jet classification by training a classifier that used the vertex finding predictions as input, as a sort of supervised attention mechanism. We found that improved vertex finding lead to improved classification. The supervised attention mechanism lead to better results compared to an identical model with un-supervised attention. The universal models (S2G and RNN) had the best performance, however the equivariance of S2G gave it a slight advantage over the RNN.

Future work may explore the application of this technique to more  complicated decays such as boosted Higgs to (bb/cc), and apply it to more realistic datasets that include full detector simulation and pileup interactions.

\section{Acknowledgments}

EG and JS are supported by the NSF-BSF Grant 2017600 and the ISF Grant 125756. This research was partially supported by the Israeli Council for Higher Education (CHE) via the Weizmann Data Science Research Center. KC is supported by the National Science Foundation under the awards ACI-1450310, OAC-1836650, and OAC-1841471 and by the Moore-Sloan data science environment at NYU. HS, NS and YL were supported in part by the European Research Council (ERC Consolidator Grant, "LiftMatch" 771136), the Israel Science Foundation (Grant No. 1830/17) and by a research grant from the Carolito Stiftung (WAIC).

\bibliography{JetGraphs}
\bibliographystyle{unsrt}

\clearpage
\appendix
\section{Hyperparameter Optimization for AVR}
\label{sec:raveoptim}

The AVR algorithm in RAVE~\cite{Waltenberger:2011zz} has three main parameters that can be adjusted by the user - 
\begin{itemize}
	\item Primary vertex significance cut
	\item Secondary vertex significance cut
	\item  minimum weight for a track to stay in a fitted vertex
\end{itemize}

\begin{figure}
	\begin{center}
		\includegraphics[]{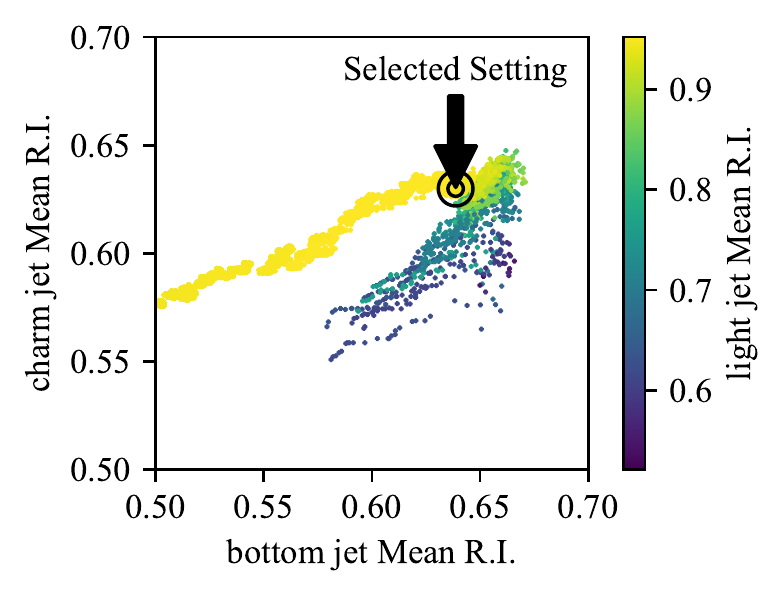}
	\end{center}
	\caption{AVR parameter scan}
	\label{fig:rave_scan}
\end{figure}

The values for there parameters were scanned in a grid between 0.1 to 10 for the significance cuts (33 equally spaced values) and between 0.1 to 0.8 for the minimum weight (10 values). For each possible value of the parameters, the mean RI was computed for each of the 3 flavors in the training dataset. The values of the b, c and light jet RI are shown in figure~\ref{fig:rave_scan}. The working point that was chosen had the highest b jet RI with a mean light jet RI above 0.95:

\begin{itemize}
	\item Primary cut: 2.5
	\item Secondary cut: 2.5
	\item  minimum weight: 0.2
\end{itemize}

\section{Model Architecture and Training Details}
\label{sec:moremodel}

\paragraph{Hyperparameter tuning and ablation studies}
The optimization of the model hyperparameters and architecture used in this paper are described in detail in the supplementary material of~\cite{serviansky2020set2graph}. Below we describe the architecture for the final optimized model.

\begin{figure}[ht]
	\begin{center}	
		\includegraphics[]{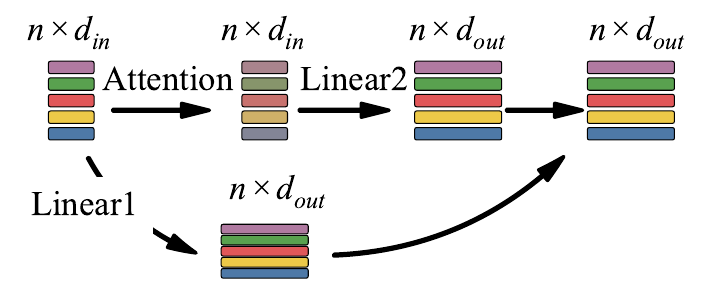}
		\caption{A single deep set layer in the $\phi$ module.}
		\label{fig:deepsetlayer}
	\end{center}
\end{figure}

\paragraph{S2G model.}
The $\phi$ component of the S2G model is composed of a sequence of deep set layers~\cite{zaheer2017deep}, each of which contain a self-attention mechanism and two linear $d_{in}\rightarrow d_{out}$ layers, in a structure shown in figure~\ref{fig:deepsetlayer}. A ReLU non-linearity is used between the layers.

The attention block in the deep set layer is a key/query attention ~\cite{ilse2018attention,vaswani2017attention}:
\begin{equation}
	\mathrm{Attention}(X) = \text{softmax} \left({\frac{\tanh{f_1(X)} \cdot f_2(X)^T}{\sqrt{d_{small}}}}\right) \cdot X
\end{equation} \label{eq:attention}

Where $X$ is the $n \times d_{in}$ input, $f_1, f_2$ are the key and query MLPs of width $d_{small} = d_{in} / 10$.

If we describe the stack of deep set layers by their output dimension $d_{out}$, the $\phi$ module layer dimensions are:

\begin{equation}
	\phi \text{ output dimensions} = (256, 256, 256, 256, 5)
\end{equation}

The edge classifier component $\psi$ takes in the $n\cdot (n-1)\times (5\cdot3)$ output of the broadcasting layer, and uses a single hidden layer MLP with output dimensions $(256, 1)$.

\paragraph{Baseline TP Classifier.}

The MLP that replaces the deep set layers has the following output sizes:

\begin{equation}
	\phi_\mathrm{TP} \text{ output dimensions} = (384,384,384,384, 5)
\end{equation}

The edge classifier component $\psi$ is identical expect its input size is now $5\cdot2$ instead of $5\cdot3$ due to the absence of the sum in the broadcasting layer. 

\paragraph{Baseline RNN}

The GRU layer output sizes are:
\begin{equation}
	\phi_\mathrm{RNN} \text{ output dimensions} = (256, 256,128, 6)
\end{equation}

Each GRU layer is bi directional. Each direction results in a hidden representation of size $d_{out}/2$, and the results are concatenated.

\paragraph{Training Hyperparameters}

We used a batch size of 2048, Adam optimizer \cite{kingma2014adam} with learning rate of $10^{-3}$. Training takes place in less than 2 hours on a single Tesla V100 GPU. The training is stopped when the validation loss stops does not decrease for 20 epochs.

\section{Jet Classification Model Architecture}
\label{sec:jetclassificationappendix}
The model, illustrated in Figure~\ref{fig:jetclassification} is composed of four components: 

\begin{itemize}
	\item Deep set network 
	\item Vertex finding module, 
	\item Graph network~\cite{battaglia2018relational}.
	\item Jet classifier MLP.
\end{itemize}

\paragraph{Deep set} The deep set network is described in~\ref{sec:moremodel}. In the classification model it has dimensions of:

\begin{equation}
	\text{Deep set output dimensions} = (126,126,126,126)
\end{equation}

The deep set creates a hidden representation for each track in the input.

\paragraph{Vertex finding module} This is either the AVR pre-computed vertex assignment, or one of the vertex finding networks. The output of this module is an edge prediction $e_{ij}$ between any two tracks in the input set.

The graph network creates a hidden representation for the tracks based on the output of the deep set and the vertex finding module, which is treated as edge features for the fully connected graph of tracks. 

The graph network is composed of a sequence of GN blocks, each with an edge update and node update MLP.

\begin{align}
	\label{eq:global}
	g^t & =\sum\limits_{i} h^t_i \\
	\label{eq:message}
	m_i^{t+1} & = \sum\limits_{j \in N(i)} E_t(h_i^t, h_j^t, e_{ij},g^t) \\
	\label{eq:update}
	h_i^{t+1} & = U_t( h_i^t, m_i^{t+1})
\end{align}
where $h_i^t$ is the $i$th node hidden representation at step $t$, $g^t$ is the global representation of the graph (sum of all node hidden representations), $E_{t}$ and $U_t$ are the edge and node update MLPs for layer $t$ of the graph network and $e_{ij}$ is the edge prediction given by the vertex finding module for the edge between node $i$ and $j$. $N(i)$ is the node neighborhood. In this model the graph is always fully connected, so the node neighborhood contains all the nodes in the graph.
The edge update MLP has linear layers with sizes:
\begin{equation}
	E_{t}~\text{dimensions} = (126\cdot3+1, 100, 20)
\end{equation}
The node update MLP has linear layers with sizes:
\begin{equation}
	U_{t}~\text{dimensions} = (126+20, 100, 126)
\end{equation}
The graph network has 3 such GN blocks.

The jet classifier MLP takes as input the sum of track hidden representations and the jet features ($p_{T}$, $\eta$, $\phi$, jet mass). It predicts if the jet is a b,c or light jet.
\begin{equation}
	\text{Jet classifier dimensions} = (126+4, 100, 50, 3)
\end{equation}

\subsection{Jet Classifier Training}

The model is trained with a batch size of 1000, Adam optimizer and a learning rate of $5\cdot10^{-4}$, and cross entropy loss. Training takes less than 2 hours on single Tesla V100 GPU. The training is stopped when the validation loss stops does not decrease for 20 epochs.

\end{document}